\documentstyle[prl,aps,twocolumn,epsf]{revtex}
\newcommand{\lapprox}{\mbox{\raisebox{-4pt}{$\,\buildrel<\over\sim\,$}}}
\newcommand{\gapprox}{\mbox{\raisebox{-4pt}{$\,\buildrel>\over\sim\,$}}}

\newcommand{\braket}[2]{\langle#1|#2\rangle}
\newcommand{\brkt}[3]{\left\langle#1\right|#2\left|#3\right\rangle}
\newcommand{\Be}{\rm{B}}
\newcommand{\HF}{\rm{HF}}
\newcommand{\tot}{\rm{tot}}
\newcommand{\UHF}{\rm{UHF}}
\newcommand{\QMC}{\rm{QMC}}
\newcommand{\GS}{\rm{GS}}
\newcommand{\exact}{\rm{exact}}
\begin{document}
\draft
\title{Wigner molecules in quantum dots}
\author{Boris Reusch,$^1$ Wolfgang H\"ausler$^2$ and Hermann Grabert$^1$}
\address{${}^1$Fakult\"at f\"ur Physik, Albert-Ludwigs-Universit\"at,
 D-79104 Freiburg, Germany\\
${}^2$I.~Institut f\"ur Theoretische Physik, Universit\"at
Hamburg, D-20355 Hamburg, Germany}
\date{Date: \today}
\maketitle
\begin{abstract}
We perform unrestricted Hartree-Fock (HF) calculations for electrons
in a parabolic quantum dot at zero magnetic field.
The crossover
from Fermi liquid to Wigner molecule behavior is studied
for up to eight electrons and various spin components $S_z$.
We compare the results with numerically exact path-integral Monte Carlo
simulations and earlier HF studies. Even in the strongly correlated regime
the symmetry breaking HF solutions provide accurate estimates
for the energies and describe the one-particle densities qualitatively.
However, the HF approximation favors the formation of a Wigner molecule
and produces azimuthal modulations of the density for even
numbers of electrons in one spatial shell.
\end{abstract}

The last decade has seen an enormous interest in
quantum dots, i.e., a small number of 2D electrons
confined in a semiconductor heterostructure \cite{jacak98}.
Experimentally, the $N$-electron states of these systems are
studied by means of far-infrared \cite{meurer92}, capacitance \cite{ashoo96}
and transport spectroscopy \cite{kouwen97}, and exhibit
features of quantization of charge and energy.
Theoretically, a whole arsenal of methods for interacting electronic systems
together with increasing computational power applies:
Exact diagonalization techniques \cite{merkt91}, density functional theory
\cite{hiros99} and quantum Monte Carlo methods \cite{harti00}.
Recently, also the very strongly correlated regime
of small electronic densities has attracted considerable interest
\cite{egger99,yanno99,reima00}.
In particular, the formation of a Wigner molecule was studied with
Quantum Monte Carlo (QMC) \cite{egger99}, Hartree-Fock (HF)\cite{yanno99}
and configuration interaction \cite{reima00} calculations.

In this paper we reconsider the Hartree-Fock approximation
focusing on the crossover to the Wigner regime. An important
practical and conceptual question to be clarified is how correlations
beyond the mean field approximation contribute to the exact energy
and whether the Wigner molecule is described reliably within HF.
In its unrestricted version, allowing for symmetry-broken solutions,
HF approaches the true ground state energy considerably better than
restricted HF which preserves the rotational symmetry
of the Hamiltonian. This is achieved, however, on expense of the
quality of the wave functions.

In the case of a strong central potential in three dimensions,
such as in real atoms, the HF approximation is known to yield useful
results both for energies and wave functions. Here, we show
by comparison with exact Monte Carlo data \cite{egger99} that even
for strong interaction {\it unrestricted} HF can give very good
estimates for the ground state energies of 2D quantum dots.
Only tiny energy differences between different spin states
cannot be resolved reliably.
On the other hand, the charge density distribution resulting
from unrestricted HF cannot quantitatively describe the
strongly correlated regime. For even number of electrons per shell,
the HF densities show effects of localization due to the
strong electron interaction.
Furthermore, within HF \cite{yanno99} this crystallization sets in
rather too early at higher densities than in the QMC study \cite{egger99}.

{\sl Model ---} We study a two-dimensional parabolic quantum dot
with $N$ electrons at zero magnetic field as it has been discussed
by many authors [5-10].
Measuring energy in units of the oscillator energy $\hbar\omega_0$ and
length in units of $l_0=\sqrt{\hbar/m^*\omega_0}$,
where $m^*$ is the effective mass, the dimensionless Hamiltonian reads
\begin{eqnarray}
   H & = & \sum_{j=1}^N \left(-\frac{1}{2}\Delta_j +
   \frac{1}{2} \bbox{r}_j^2 \right) +
   \sum_{i<j=1}^N \frac{\lambda}{|\bbox{r}_i-\bbox{r}_j|}\nonumber \\
  &  \equiv &\sum_{i=1}^N h_i + \sum_{i<j=1}^N w_{ij} \;.
\end{eqnarray}
Here we have introduced the dimensionless coupling constant
$\lambda=l_0/a^*_{\Be}=e^2/\kappa l_0\hbar\omega_0$ with
the effective Bohr radius $a^*_{\Be}$ and the dielectric constant $\kappa$.
For example $\lambda\!=\!2$ corresponds to $\hbar\omega_0\!\approx\!3$meV
for a GaAs quantum dot.
Since this Hamiltonian is rotationally invariant and spin independent,
the exact eigenfunctions can be chosen as simultaneous eigenfunctions
of the total angular momentum $L^{\tot}_z$, the total spin $\bbox{S}^2_{\tot}$
and its $z$-component $S_z^{\tot}$.
These eigenfunctions and corresponding densities are then rotationally
invariant.

{\sl Method ---} HF theory consists in approximating the many-particle
wave function by an optimal single Slater determinant
\begin{equation}
 \Psi^{\HF} = \frac{1}{\sqrt{N!}} \: \det \left( \varphi_i(\bbox{r}_j)\right)_{1\le i,j\le N}\;.
\end{equation}
This HF Slater determinant is build up by single particle orbitals
$\varphi_i(\bbox{r})$, which are expanded in the angular momentum basis
of the 2D harmonic oscillator (Fock-Darwin states)
\begin{equation}
\label{basis}
\braket{\bbox{r}}{ n M} = \sqrt{\frac{n!}{\pi(n+|M|)!}}\;e^{iM\varphi}\;r^{|M|}
\:{\cal{L}}^{|M|}_n\! (r^2)\: e^{-r^2/2} \;.
\end{equation}
Here, $n$ and $M$ are the radial and angular quantum numbers, and
${\cal{L}}^{|M|}_{n}$ is a Laguerre polynomial.
In the unrestricted HF approximation an orbital has the expansion
\begin{equation}
\label{orbex}
\varphi_i(\bbox{r}) = \sum_{{n=0,\infty \atop M=-\infty,\infty}} u_{nM}^i \braket{\bbox{r}}{n M \sigma_i} \;,
\end{equation}
where $\sigma_i\!=\!\pm\frac{1}{2}$ is the fixed electron spin of
the $i$-th orbital.
This orbital is no longer an eigenfunction of the one particle angular
momentum, therefore the HF Slater determinant is in general not an eigenfunction
of $L^{\tot}_z$, but a deformed, symmetry broken solution \cite{rhf1}.
Similarly, the HF solution is in general not an eigenfunction
of the total spin but only of its $z$-component with eigenvalue
$S_z^{\tot} \equiv S_z\!=\!(N_\uparrow- N_\downarrow)/2\!=\!\sum_i\sigma_i$.

Minimizing the HF energy $E^{\HF}\! = \!\brkt{\Psi^{\HF}}{H}{\Psi^{\HF}}$
and imposing orthonormality between the HF orbitals
yields the HF equations for the expansion coefficients $u_{nM}^i$
\begin{equation}
 \label{hfglg}
 \sum_{\alpha} \big\{\brkt{\gamma}{h}{\alpha}+\sum_{\beta\beta'}(\gamma\beta|w|\alpha\beta')\rho_{\beta'\beta}\big\} u^k_{\alpha}=\varepsilon_k u^k_{\gamma} \;.
\end{equation}
Here, greek indices abbreviate the quantum numbers of (\ref{basis}),
for example $\alpha \equiv (n,M,\sigma)$.
The density matrix is defined by
$\rho_{\alpha\alpha'} = \sum_{i=1}^{N} u^i_{\alpha}{u^i_{\alpha'}}^*$,
$\brkt{\gamma}{h}{\alpha}$ is the one particle matrix element and
$(\alpha\alpha'|w|\beta\beta')=\brkt{\alpha\alpha'}{w}{\beta\beta'}-\brkt{\alpha\alpha'}{w}{\beta'\beta}$
denotes the antisymmetrized Coulomb matrix element,
which we calculated analytically by transformation
into relative and center of mass coordinates \cite{reusc98}.
The nonlinear selfconsistent eigenvalue problem (\ref{hfglg})
for the expansion coefficients $u^i_{\alpha}$ is solved by iterative
diagonalization starting from an initial guess for the density matrix.
Several starting guesses have to be chosen in order to avoid
local minima. The true minimum can also be identified by its one particle
density,
\begin{equation}
\label{dens}
n^{\HF}(\bbox r) = \sum_{i=1}^N |\varphi_i(\bbox{r})|^2 \;,
\end{equation}
which in the strong coupling limit mirrors the geometry of classical
electrostatic point-charges as we discuss below.

In our calculations we used up to 55 Fock-Darwin states for each
spin direction. We present results for various sets of ${N,\lambda,S_z}$
from $N=2$ to $8$ particles covering the
whole range of interaction strengths from $\lambda=2$ to $10$.

{\sl HF energies and spin ---} The most important properties of the
HF ground state energies can already be seen for $N=2$ (quantum dot
Helium). Here one knows the exact solution from diagonalization
of the Hamiltonian for the relative motion \cite{merkt91}.
We have reproduced these results in Table~\ref{table1}.
The true ground state is always a singlet
and the energy gap to the triplet vanishes slowly as $\lambda\to\infty$.
In contrast, the HF approximation finds a triplet ground state
for $\lambda \gapprox 5$,
and the energies of singlet and triplet converge more rapidly \cite{rhf2}.
The (absolute and relative) error is largest in the case of
$S_z=0$ and $\lambda \approx 2$.
In the polarized case the error of HF increases monotonically with $\lambda$.
These features of the HF energies can be seen for all $N$.

For $N > 2$ we compare with the path integral QMC
data of Ref.~\cite{egger99}, which have been obtained for the very
low temperature $T\!=\!0.1\hbar\omega_0/k_{\Be}$.
The HF ground state energies are always above the QMC energies
for all particle numbers investigated so that the QMC data may serve
effectively as zero temperature reference points.
In the case of $N=5$ electrons, QMC gives a ground state spin of $1/2$.
In Fig.~\ref{diff5} we plot the energy difference of the different spin
states with respect to the true ground state, $\Delta E=E_S-E_{1/2}^{\QMC}$,
for various spin states as a function of the coupling strength $\lambda$.
Again we see that $\Delta E$ is largest for $\lambda \approx 2$
and smallest spin. It remains constant for $\lambda \gapprox 4$,
resulting in a very low relative error of only $1-2$\% in the low density
regime. The HF energy is very accurate in the case of $S_z=5/2$ which becomes
the HF ground state for $\lambda \gapprox 4$. This unphysically
high ground state spin is due to the exchange term in the HF energy,
which lowers the energy only for parallel spins.
Therefore the spin ordering of the HF energies is just reversed.

For $N=8$ we give the HF energies in Table~\ref{table2}. The QMC predicts
a transition of the ground state spin from $S=1$ to $S=2$
for $\lambda \gapprox 4$. Again HF finds the wrong spin ordering in
the correlated regime.
The relative error is largest for the unpolarized
states. The convergence of the HF energies for different $S_z$ for
$\lambda \gapprox 6$ can be understood in the classical picture of
localized electrons without overlap and therefore no spin sensitivity.

{\sl HF densities ---} Next we study the HF one particle densities
(\ref{dens}), where the crossover to the Wigner molecule
is discernible. However,
one has to keep in mind that the deformation and structure
of the HF densities arise from the symmetry violating mean field
and therefore may be artificial (whereas the HF {\it energies}
are a true upper bounds for the ground state energies). Second
we want to point out that a symmetry breaking {\it Slater determinant} does
not necessarily mean that its corresponding {\it density} displays
a molecule-like structure.
In Fig.~\ref{l6} we display the densities for $N=2$ to $5$
for strong coupling $\lambda = 6$ and maximal spin $S_z = N/2$
(From $\lambda\gapprox 6$ the HF densities are essentially the same
for all spins.).
In the case of $2$ and $4$ electrons
they show quite distinct azimuthal maxima. This is not the case
for the odd electron numbers $3$ and $5$: Surprisingly the HF densities,
though belonging to a deformed Slater determinant, seem to be
rotationally symmetric.

In order to understand this even-odd effect consider an exact
spin-polarized $N$-electron wave function $\Psi_N$ for the Wigner
molecule case. An even number of electrons in one spatial shell
carries a {\it nonzero} angular momentum $\pm \hbar N/2$ \cite{leggett}.
In this fashion the HF wavefunctions with modulated density ($N=2,4$)
can be interpreted as standing waves from a superposition of opposite
angular momenta \cite{cdw}.

However the HF densities display the right classical filling
for the spatial shells:
From the maxima in the densities one can also read off the
Brueckner parameter $r_s$, which is defined as the the nearest-neighbor
distance in units of $a^*_{\Be}$.
These values $r_s$ agree well with those obtained by QMC
and from a model of classical point-charges.
In Fig.~\ref{n7} we show a 7-electron Wigner molecule
for two interaction strengths. For $\lambda = 2$
the HF density is already 6-fold modulated with a central electron.
Confronting this with the exact density in Fig.~1 of
Ref.~\cite{egger99}, we notice that the crystallization
(as suggested by the HF densities) occurs too early.
This explains why the authors of Ref.~\cite{yanno99} observe
Wigner crystallization for higher densities than in the exact QMC study
\cite{egger99}.
For higher $\lambda$ the molecule with two spatial shells
becomes more and more distinct. The maxima
agree very well with the classical formula $r_s^3 = \lambda^4 (2.25 +
1/\sqrt{3})$ for the 6-fold geometry.

Finally we show in Fig.~\ref{n6} the crystallization for $N=6$ and $S_z=0$.
This unpolarized case depicts a crossover from a 6-fold arrangement
[Fig.~\ref{n6}(a)] to a 5-fold geometry [Fig.~\ref{n6}(c) and (d)].
The 6-electron molecule was also studied by Reimann et~al.
\cite{reima00}, who found by means of configuration interaction
calculations that the true ground state was unpolarized with a 6-fold symmetry
up to at least $\lambda \approx 3.5$.
Within HF the $S_z=0$ state acquires a 5-fold symmetry already for
$\lambda \ge 2.85$. In this range the HF density is distorted
[$\lambda = 4$ in Fig.~\ref{n6}(c)] and then again apparently round
for higher lambda [$\lambda = 6$ in Fig.~\ref{n6}(d)] with a central maximum.
Fig.~\ref{n6}(b) shows the 6-fold isomer for $\lambda = 4$, which is
by $0.33$ higher in energy than the ground state in Fig.~\ref{n6}(c).
In contrast the spin-polarized state exhibits 5-fold symmetry
throughout the whole parameter range \cite{comment}.

In conclusion we have shown in how far an unrestricted HF description
of the Wigner molecule in quantum dots is reliable.
The HF energies are good estimates of the true ground state energies
especially for spin-polarized states. The energy differences
for different spin-states in the strong interacting regime cannot
be resolved properly.
We find deformed HF densities in the regime of intermediate
interaction up to $\lambda \lapprox 4$.
For strong correlation the densities are azimuthally modulated
for an even number of electrons in a shell and round for an
odd number per shell. The onset of this modulation is enhanced
within HF, which leads to overestimate the value of the critical
density for the crossover to the Wigner molecule.
However the HF densities mirror the classical
filling scheme with the electrons arranged in spatial shells.

We thank R.~Bl{\"u}mel, R.~Egger, C.~Stafford, and T.~Vorrath
for useful discussions.
This research has been supported by the SFB 276 of the Deutsche
Forschungsgemeinschaft (Bonn)

\begin{table}
\caption{\label{table1} Lowest energies of exact diagonalization and HF
for $N=2$ and various $S_z$ and coupling strengths $\lambda$. Relative
error $(E_S^{\exact}-E_S^{\exact})/E_S^{\exact}$ in $\%$.}
\begin{tabular}{|ccccc|}
$\lambda$ & $S_z$ & $E^{\exact}$&$E^{\HF}$&rel.~err.\\\hline
2 & 1 &  4.142 & 4.168 & 0.7\\
2 & 0 &  3.721 & 4.034 & 8.3\\\hline
4 & 1 &  5.119 & 5.189 & 1.4\\
4 & 0 &  4.848 & 5.182 & 6.8\\\hline
6 & 1 &  5.990 & 6.096 & 1.8\\
6 & 0 &  5.784 & 6.107 & 5.7\\\hline
8 & 1 &  6.787 & 6.919 & 1.9\\
8 & 0 &  6.618 & 6.930 & 4.7\\\hline
10& 1 &  7.528 & 7.679 & 2.0\\
10& 0 &  7.384 & 7.686 & 4.2\\
\end{tabular}
\begin{figure}[htb]
\centerline{\epsfxsize=9cm \epsfysize=7cm \epsffile{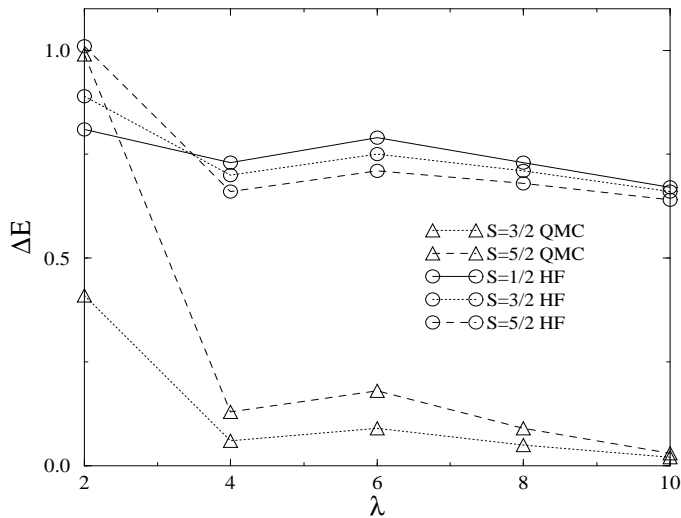}}
\caption{$N=5$. Absolute energy differences to QMC ground state ($S=1/2$),
 $\Delta E=E_S-E_{\GS}^{\QMC}$, for various spins vs. coupling
constant $\lambda$}
\label{diff5}
\end{figure}
\begin{figure}[htb]
\centerline{\epsfxsize=9cm \epsfysize=9cm \epsffile{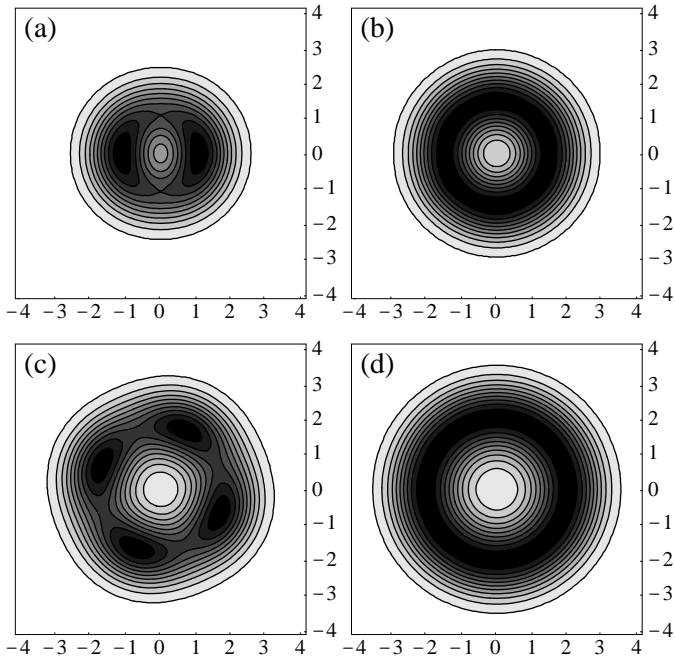}}
\caption{Shadowed contour-plot of the HF density $n^{\HF}$ for $\lambda=6$,
$S_z=N/2$ and different electron numbers. Contours lie at integral multiples
of 0.1 times the maximal density. (a) $N=2$, (b) $N=3$, (c) $N=4$, (d) $N=5$.}
\label{l6}
\end{figure}
\begin{figure}[htb]
\centerline{\epsfxsize=9cm \epsfysize=4.5cm \epsffile{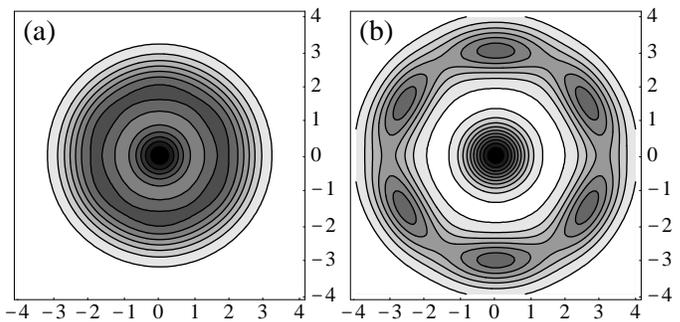}}
\caption{7-electron Wigner molecule. HF density $n^{\HF}$ for (a) $\lambda=2$ $S_z=1/2$, (b) $\lambda=10$ $S_z=7/2$.}
\label{n7}
\end{figure}
\begin{figure}[htb]
\centerline{\epsfxsize=9cm \epsfysize=9cm \epsffile{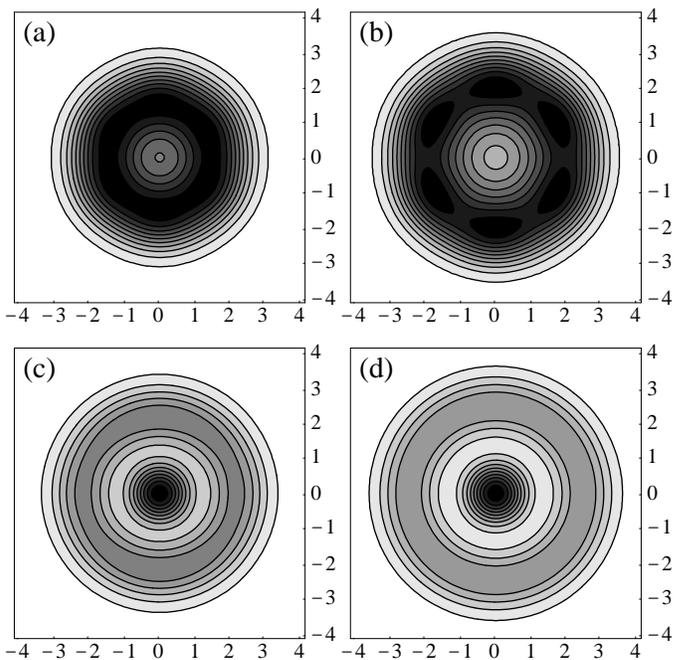}}
\caption{$N=6$ $S_z=0$. Rearrangement from a 6-fold [(a) $\lambda=2$] to a
(deformed) 5-fold geometry [(c) $\lambda=4$, $E=41.509$, (d) $\lambda=6$]. In
(b) 6-fold isomer at $\lambda=4$ with energy $E^*=41.838$.}
\label{n6}
\end{figure}
\end{table}
\begin{table}
\caption{\label{table2} Energies of QMC vs. HF for $N=8$.
Bracketed numbers denote statistical errors of QMC. Relative error
$(E_S^{\UHF}-E_S^{\QMC})/E_S^{\QMC}$ in $\%$.}
\begin{tabular}{|ccccc|}
 $\lambda$ & $S_z$ & $E^{\QMC}$&$E^{\HF}$&rel.~err.\\ \hline
 2 & 4   & 48.3(2)  & 48.534  & 0.5 \\
 2 & 3   & 47.4(3)  & 48.336  & 2.0 \\
 2 & 2   & 46.9(3)  & 48.243  & 2.9 \\
 2 & 1   & 46.5(2)  & 48.132  & 3.5 \\\hline
 4 & 4   & 69.2(1)  & 69.735  & 0.8 \\
 4 & 3   & 68.5(2)  & 69.783  & 1.9 \\
 4 & 2   & 68.3(2)  & 69.826  & 2.2 \\\hline
 6 & 4   & 86.92(6) & 87.957  & 1.2 \\
 6 & 3   & 86.82(5) & 87.999  & 1.4 \\
 6 & 2   & 86.74(4) & 88.039  & 1.5 \\\hline
 8 & 4   & 103.26(5)& 104.492 & 1.2 \\
 8 & 3   & 103.19(4)& 104.520 & 1.3 \\
 8 & 2   & 103.08(4)& 104.547 & 1.4 \\
\end{tabular}
\end{table}

\begin{thebibliography}{12}
\bibitem{rhf2}
In the {\it restricted} HF the energies for different $S_z$ do not
merge, resulting in a considerably larger error for the unpolarized states.
\bibitem{cdw}
A corresponding effect for weak interaction was observed in
the density-functional study of Ref.~\cite{hiros99}.
\end{thebibliography}
\end{document}